\documentclass[]{aa}
\usepackage{graphicx}
\usepackage{psfig}
\usepackage{amssymb}
\usepackage{verbatim}
\usepackage{natbib}
\usepackage{rotate}
\usepackage{lscape}
\usepackage{aalongtable}
\usepackage{supertabular}
\usepackage{mathbbol}

\bibpunct{(}{)}{;}{a}{}{,}

\newcommand{\appropto}{\mathrel{\vcenter{
  \offinterlineskip\halign{\hfil$##$\cr
    \propto\cr\noalign{\kern2pt}\sim\cr\noalign{\kern-2pt}}}}}

\begin{document}      
	      
\title{Applying full polarization A-Projection to very wide field of view instruments: An imager for LOFAR}

\subtitle{}
\author{C. Tasse\inst{1,2,3}, B. van der
  Tol\inst{4}, J. van Zwieten\inst{5}, Ger van Diepen\inst{5}, \and S. Bhatnagar\inst{6}}

\institute{GEPI, Observatoire de Paris, CNRS, Universit\'e Paris Diderot,
5 place Jules Janssen, 92190 Meudon, France
\and
Department of Physics \& Electronics, Rhodes University, PO Box 94,
Grahamstown, 6140, South Africa
\and
SKA South Africa, 3rd Floor, The Park, Park Road, Pinelands, 7405, South Africa
\and 
Sterrewacht Leiden, PO Box 9513, 2300 RA, Leiden, The Netherlands
\and
Netherlands Institute for Radio Astronomy (ASTRON), Postbus 2,
7990 AA Dwingeloo, The Netherlands
\and
National Radio Astronomy Observatory, Socorro, NM 87801, USA}
\date{Received <date> / Accepted <date>}

\abstract{The aimed high sensitivities and large fields of view of the
  new generation of interferometers impose to reach high dynamic range
  of order $\sim$1:$10^6$ to $1$:$10^8$ in the case of the Square
  Kilometer Array. The main problem is the calibration and
  correction of the Direction Dependent Effects (DDE) that can affect
  the electro-magnetic field (antenna beams, ionosphere, Faraday
  rotation, etc.). As shown earlier the A-Projection is a fast and
  accurate algorithm that can potentially correct for any given DDE
  in the imaging step. With its very wide field
  of view, low operating frequency ($\sim30-250$ MHz), long baselines,
  and complex station-dependent beam patterns, the Low Frequency Array
  (LOFAR) is certainly the most complex SKA precursor. In this paper
  we present a few implementations of A-Projection applied to LOFAR
  that can deal with non-unitary station beams and non-diagonal
  Mueller matrices. The algorithm is designed to correct for all the
  DDE, including individual antenna, projection of the dipoles on the
  sky, beam forming and ionospheric effects.  We describe a few
  important algorithmic optimizations related to LOFAR's architecture
  allowing us to build a fast imager. Based on simulated datasets we
  show that A-Projection can give dramatic dynamic range improvement
  for both phased array beams and ionospheric effects. We will use
  this algorithm for the construction of the deepest extragalactic
  surveys, comprising hundreds of days of integration.}

\authorrunning{C. Tasse}

\titlerunning{Applying full polarization A-Projection to very wide fields of view instruments}
   \maketitle

\section{Introduction: LOFAR and the direction dependent effects}

With the building or development of many large radio telescopes (LOFAR, EVLA,
ASKAP, MeerKAT, MWA, SKA, e-Merlin), radio astronomy is undergoing a period of
rapid development. New issues arise with the development of these new types of
interferometer, and the approximations 
applicable to the older generation of instruments are
not valid anymore. Specifically, they have wide fields of
view and will be seriously affected by the Direction Dependent Effects
(DDE).
Dealing with the DDE
in the framework of calibration and imaging represents an unavoidable
challenge, on both the theoretical, numerical and technical aspects of
the problem \citep[see][for a detailed review of the problems
associated with calibration and wide field imaging in the presence of DDE]{Bhatnagar09}.


This is particularly true for the Low Frequency Array (LOFAR). It is
an instrument that observes in a mostly unexplored frequency range
($\nu\lesssim240$ MHz), and will be one of the largest radio
telescopes ever built in terms of collecting area. LOFAR's design is
built on a combination of phased array and interferometer (see
\citet{deVos09} for a description of the LOFAR system). It is made of $40$
stations in the Netherlands, and 8 international stations (5 in
Germany, 1 in France, England, and Sweden). The High Band Antenna
stations (110-240 MHz, HBA hereafter) are made of 24 to 96 {\it tiles} of
$4\times 4$ antenna coherently summed, while the Low Band Antenna
(10-80 MHz, LBA) are clustered in groups of 96 elements. At the
station level, the signals from the individual antennas or tiles (in
the cases of LBA and HBA respectively) are phased and summed by the
{\it beamformer}. This step
amounts to forming a virtual antenna pointing at the targeted field
location. The data is transported from the various stations of the
LOFAR array to the correlator. The whole process and the pipeline
architecture have been described in more details in
\citet{Heald10}. LOFAR is affected by many complex
baseline-time-frequency$^{\ref{foot:bl}}$ dependent DDE, consisting mainly of the
antenna/station beams and the ionosphere, which varies on angular
scales of degrees and time scales of $\sim10$ minutes and $\sim30$
seconds respectively. We currently have models of both the high-band
and low-band station beams (HBA and LBA respectively).


As shown in \citet{Bhatnagar08} A-Projection allows to estimate sky
images, taking into account all the possible complicated effects
associated to the DDE \citep[see also][in the context of the
Murchison Widefield Array and forward modeling]{Bernardi11,Mitchell12,Sullivan12}. However contrarily to dishes-based interferometers, where
the beam shape and polarization angle are affected by pointing errors
and rotated on the sky by the parallactic angle (depending on
the dish mount), LOFAR is based on phased arrays that have very wide
fields of view (up to $\sim12$ degrees), non-trivial and quickly varying beams, thereby driving complicated
polarization effects. Technically speaking, the very wide fields of view instruments
that aim to reach high dynamic range, have to deal with
baseline-dependent non-diagonal Mueller matrices (see
Sec. \ref{sec:AProj} for a detailed discussion). For the VLA
implementation, due to the
approximate {\it Unitarity} of VLA beams, it was sufficient for A-Projection to take into account the
diagonal terms of the Mueller matrices to demonstrate corrections for
instrumental polarization. That is not possible for LOFAR that has
heavily non-diagonal baseline-associated
Mueller matrices, and all $4\times 4$ Mueller terms have to be
taken into account.

We show in this paper that the scheme described in \citet{Bhatnagar08}
can indeed deal with the heavily non-unitary beams associated with the
very wide field of view of phased arrays. Our imaging algorithm could take
as input any model or calibration solution or ionosphere phase
screen. In Sec. \ref{sec:AProj} we describe the issues related with the
usage of phased arrays in interferometers, and focus on the
LOFAR-related issues {\it i.e.} the polarization aspects and
baseline dependence of the DDE. We describe a
few important algorithmic optimizations related to LOFAR's
architecture allowing us to build a fast imager\footnote{Our software ({\it awimager}) is built 
on the Casa imager implementation.}. In Sec. \ref{sec:impl} we describe the
A-Projection algorithm first presented in \citet{Bhatnagar08}, and
detail the various implementations and optimizations we have found to
make it reasonably fast in the case of LOFAR. We present the results
in Sec. \ref{sec:sim} and show that beam and ionosphere corrections
can both be performed at high accuracy. We summarize and discuss the next developments in
Sec. \ref{sec:concl}.

\section{Polarization effects associated with very wide fields of view
  interferometers}
\label{sec:AProj}

\begin{figure*}[ht!]
\begin{center}
\hspace*{-1.3cm}
\includegraphics[width=13cm]{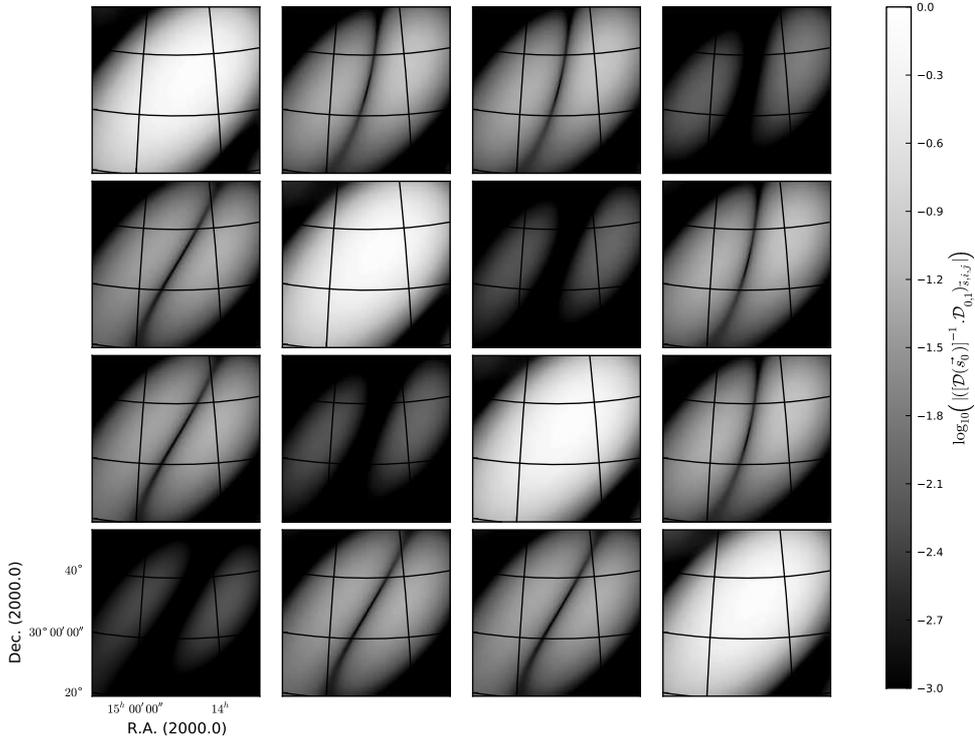}
\caption{\label{fig:beams} LOFAR stations are phased arrays characterized
  by a large field of view. The X/Y polarimetric measurements are therefore non
  trivial standard compared to a radio telescope using dishes: we have
  to take into account the projection of the dipoles that are
  generally non-orthogonal on the sky and the fact that this angle
  varies across the field of view. This figure shows the Mueller
  matrix corresponding to baseline (01) in a given time and frequency
  slot, normalized by the Mueller matrix at the phase center (see text). Each pixel in the plot $(i,j)$ shows the
  amplitude of the $(i,j)$ Mueller matrix element in a certain direction $s$ using a logarithmic
  scale. Even in this normalized version, the off-diagonal Mueller
  terms are as high as 10\% and cannot be neglected.}
\end{center}
\end{figure*}

In this section we describe the polarizations effects associated with the complex structure of the DDE
inherent to the usage of phased arrays that have very wide fields of
view and non-diagonal
baseline-dependent Mueller matrices (or non-unitary Jones matrices in the
case of interferometers having similar antennas, see Fig. \ref{fig:beams} and the discussion in Sec. \ref{sec:AOperator}). 

With its long baselines, large fractional bandwidth, very wide field of view and station-dependent
effective Jones matrices (beam, ionosphere, Faraday rotation), the Mueller matrices to be
considered are not only non-diagonal, but are also
baseline-dependent. In order to highlight some of the main
complications associated with very wide fields of view instruments, in Sec.
\ref{sec:AOperator} we describe in detail the structure of the linear
operator introduced by \citet{Bhatnagar08}. We propose in Sec. \ref{sec:poleffects} a method to approximately correct for the associated effects corrupting the image plane
 \citep[see also][for other examples of the use of
  linear operator in the context of image synthesis and
  deconvolution]{Rau11}.


\subsection{Description of the baseline-dependence, DDE and
  polarization effects}
\label{sec:AOperator}

For convenience, in this section and throughout the paper, we do not show the sky term
$\sqrt{1-l^2-m^2}$ that usually divide the sky to account for the
projection of the celestial sphere on the plane, as this has no
influence on the results. The DDE
below are baseline-dependent, as it is the case for LOFAR, and the Measurement
Equation formalism 
can properly model those effects \citep[for extensive discussions on validity and
  limitations of the Measurement Equation
  see][and see Appendix \ref{sec:ME} for a short
  review of our needs]{Hamaker96,Oleg11}. If $V^{meas}_{pq}$ is the set of 4-polarization measurement (XX, XY, YX, YY), $G_p$ is the direction independent Jones matrix
of antenna-$p$, then the corrected visibilities on baseline $pq$ are
$V^{corr}_{pq}=[G_{t\nu,p}]^{-1}.V^{meas}_{pq}.[G^{H}_{t\nu,q}]^{-1}$, and we can
write:

\begin{equation}
\label{eq:ME_Im}
\begin{array}
{lcl}
\textrm{Vec}(V^{corr}_{pq})    &=&\int\limits_S (D^{t\nu,*}_{q,\vec{s}} \otimes D^{t\nu}_{p,\vec{s}}).\textrm{Vec}(I_{\vec{s}}) \\ 
 & & .\exp{\left(-2 i\pi \phi(u,v,w,\vec{s})\right)} d\vec{s} \\ 
\end{array}
\end{equation}

\noindent where $I$ is the 4-polarization sky, $\otimes$ is the Kronecker product, $\textrm{Vec}$ is
the operator$^{\ref{foot:vec}}$ that transforms a 2$\times$2 matrix into a dimension 4
vector, and
$\phi(u,v,w,\vec{s})=u.l+v.m+w.(\sqrt{1-l^2-m^2}-1)$
models the product of the effects of correlator, sky brightness and
array geometry. The matrix $D^{t\nu*}_{q,\vec{s}} \otimes D^{t\nu}_{p,\vec{s}}$ is a
$4\times 4$ matrix, and throughout the text we refer to it as the {\it
  Mueller} matrix\footnote{This is not completely true as
traditionally the Mueller matrix multiplies an (I, Q, U, V) vector and not
an (XX, XY, YX, YY) correlation vector.}. We can also write Eq. \ref{eq:ME_Im} in terms a series of linear transformations:

\begin{equation}
\label{eq:block}
\begin{array}
{lcl}
V^{t,\nu}_{pq}    &=& W^{t,\nu}_{pq}.S^{t,\nu}_{pq}.F.\mathcal{D}^{t,\nu}_{pq}.I \\
\end{array}
\end{equation}

\noindent where $V^{t,\nu}_{pq}$ are the $4N^{t,\nu}_{pq}$ visibility
measurement points in the time-frequency block in which the direction
dependent effects are assumed to be constant. If $N_{pix}$ is the
number of pixels in the sky model, the {\it true} sky image
vector $I$ has a size of $4N_{pix}$ and contains the full polarization
information $(XX_x,XY_x,YX_x,XY_x)$ on the $x^{th}$ pixel at the $4x$
position. $\mathcal{D}^{t,\nu}_{pq}$ contains the direction dependent effects, and
is a $(4N_{pix})\times(4N_{pix})$ block diagonal matrix. On a given
baseline (p,q), each of its $4\times4$ block is the
$\mathcal{D}^{t,\nu}_{pq}(\vec{s}_x)=D^*_q(\vec{s}_x)\otimes D_p(\vec{s}_x)$
Mueller matrix evaluated at the location of the $x^{th}$ pixel. $F$ is
the Fourier transform operator of $(4N_{pix})\times(4N_{pix})$. Each
of its $(4\times4)$ block is a scalar matrix, the scalar being the kernel of the Fourier basis $\exp{\left(-2 i\pi
  \phi(u,v,w,\vec{s})\right)}$. The matrix $S^{t,\nu}_{pq}$ is
the uv-plane sampling function for that visibility of size
$4N^{t,\nu}_{pq}\times 4N_{pix}$, and $W^{t,\nu}_{pq}$ is the diagonal
$4N^{t,\nu}_{pq}\times 4N^{t,\nu}_{pq}$ matrix containing the weights
associated with the $4N^{t,\nu}_{pq}$
visibilities.

\begin{figure*}[ht!]
\begin{center}
\hspace*{-1.3cm}
\includegraphics[width=13cm]{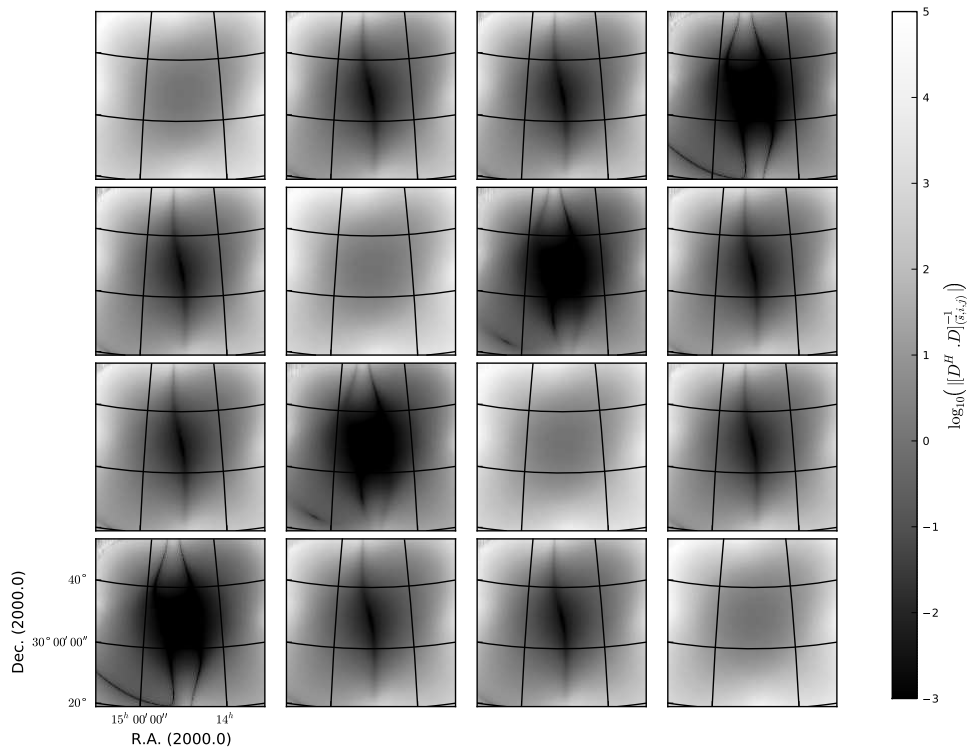}
\caption{\label{fig:beams_inv} This figure shows the correction
  $(\overline{\mathcal{D}_0^H.\mathcal{D}_0})^{-1}$ that can be applied to the
  image before the minor cycle. This is a first order correction for
  the complicated phased array beam, that depends on time, frequency
  and baseline.}
\end{center}
\end{figure*}

We show in Fig. \ref{fig:beams} that the Mueller matrix is non-diagonal
for LOFAR. The various panels show the amplitude of the direction
dependent Mueller matrix using our beam model only (therefore it does
not include Faraday rotation and ionosphere), for the baseline
$(p,q)=(0,1)$ in a given time and frequency slot. In order to minimize
the off-diagonal elements, we have computed
$\mathcal{D}_{0}(\vec{s})=[\mathcal{D}(\vec{s_0})]^{-1}.\mathcal{D}(\vec{s})$
in the direction $\vec{s}$, where $\vec{s_0}$ is the phase center
direction. Intuitively, the normalization of the Mueller matrix by
$\mathcal{D}(\vec{s_0})$ makes the projected dipoles on the sky
orthogonal at the phase center (off-diagonal elements are zero there),
while this gets less true as we go further from the targeted
location. The off-diagonal elements can be as high as $\sim10\%$ a few
degrees from the pointed location and contrarily to most
interferometers they cannot be neglected in the case of LOFAR.

We are generally interested in using the total set of visibilities over baselines, time
and frequencies, having $4N_{V}$ points. We can write:

\begin{equation}
\label{eq:SE}
\begin{array}
{lcl}
V    &=& A.I +\epsilon \\
\end{array}
\end{equation}

\noindent where $\epsilon$ is noise, and A is a $(4N_{V})\times
(4N_{pix})$ matrix, made of the
$W^{t,\nu}_{pq}.S^{t,\nu}_{pq}.F.\mathcal{D}^{t,\nu}_{pq}$ on top of
each other (each have dimension $4N_{block}\times 4N_{pix}$, where
$N_{block}$ is the total number of time-frequency blocks). We can also
write $\mathcal{A}=\mathcal{W}\mathcal{S}\mathcal{F}\mathcal{D}$,
where $\mathcal{D}$ has size $4N_{blocks}N_{pix}\times N_{pix}$, and
has all the $\mathcal{D}^{t,\nu}_{pq}$ on top of each
other. $\mathcal{F}$ is the block-diagonal Fourier transform operator
with size $4N_{blocks}N_{pix}\times 4N_{blocks}N_{pix}$, with all of
its blocks $4N_{pix}\times 4N_{pix}$ being equal to the matrix $F$
appearing in Eq. \ref{eq:block}. $\mathcal{S}$ and $\mathcal{W}$ are
the sampling and weights matrix of sizes $(4N_{V})\times
(4N_{blocks}N_{pix})$ and $(4N_{V})\times (4N_{V})$ respectively.
The transformation of Eq. \ref{eq:SE} is {\it exact}. Estimating a sky
from a sparsely sampled measured set of visibilities is less
trivial.

\subsection{Estimating a sky image: polarization effects}
\label{sec:poleffects}

There are different ways to solve for $I$ from the set of
visibilities, and reversing Eq. \ref{eq:SE} relies on the linearity of
the sky term in the measurement equation. As mentioned above our
deconvolution scheme uses A-Projection. This generalization of
CS-CLEAN is now better understood in the framework of compressed
sensing, which include other new techniques \citep[see][for a
  review]{McEwen11}. Compressed sensing provides new ways to estimate
the sky $I$ for the set of visibilities. In this section we describe a method to
approximatly correct the polarisation effects in the image
plane. In order to highlight
the issues associated with polarization and baseline-dependence of the
DDE, we simply write the sky least square solution $\hat{I}$ as the
pseudo inverse:

\begin{equation}
\begin{array}
{lcl}
\vspace{2mm}
\hat{I}    &=& \left(\mathcal{A}^H.\mathcal{A}\right)^{-1}.\mathcal{A}^H.V
\end{array}
\end{equation}

\noindent where the term $\mathcal{A}^H.V$ is the $4N_{pix}$ dirty image,
and $(\mathcal{A}^H.\mathcal{A})^{-1}$ is the $(4N_{pix})\times (4N_{pix})$ image plane
deconvolution matrix. Its structure is rather complicated and its size
makes its estimate prohibitive.

In the simple case of a matrix $\mathcal{D}$ being unity (no
DDE), we can see that each $4\times4$ block number $(x,y)$ of the matrix
$\mathcal{A}^H.\mathcal{A}$ is the instrumental response to a source
centered at the location of the $x^{th}$ pixel, evaluated at the
$y^{th}$ pixel. Therefore, computing $(\mathcal{A}^H.\mathcal{A})^{-1}$ would
involve estimating point spread function (PSF) centered at the location of each pixel and
inverting a $4N_{pix}\times 4N_{pix}$ matrix. In the presence of non-trivial baseline-dependent DDE involving
non-diagonal Mueller matrices, the problem becomes more complex. 
However, we show below that under some assumptions, the operator
$(\mathcal{A}^H.\mathcal{A})^{-1}$ (sometimes called the {\it
deconvolution matrix}) affected by DDE is decomposable in a classical
deconvolution matrix (containing information on the PSF), and a simple
correction performed separately on each pixel.

Following the notation introduced above, we can write
$\mathcal{A}^H.\mathcal{A}=\mathcal{D}^H\mathcal{P}\mathcal{D}$ as a $4N_{pix}\times
4N_{pix}$ matrix, with $\mathcal{P}\equiv\mathcal{F}^H\mathcal{S}^H\mathcal{W}^H
\mathcal{W}\mathcal{S}\mathcal{F}$ of size $4N_{blocks}N_{pix}\times
4N_{blocks}N_{pix}$. This later matrix is block diagonal, and each of its $4N_{pix}\times
4N_{pix}$ block describes the PSF of the instrument for a given
baseline-time-frequency block. Their $4\times4$ $xy$-block are scalar matrices, the scalar $p_{t,\nu,pq}(x,y)$ being the response of the
instrument evaluated at the $y^{th}$ pixel to a source being centered
at the $x^{th}$ pixel in the given baseline-time-frequency block. We
then have:

\begin{equation}
\label{eq:aproj}
\begin{array}
{lcl}
\vspace{2mm}
[\mathcal{A}^H.\mathcal{A}](x,y) &=&
\displaystyle\sum\limits_{t,\nu,pq}p_{t,\nu,pq}(x,y)\mathcal{D}^H_{t,\nu,pq}(y)\mathcal{D}_{t,\nu,pq}(x)\\
\end{array}
\end{equation}

It is virtually impossible to compute this matrix, and this
illustrates the difficulty of doing an image plane deconvolution in
the presence of time-frequency-baseline dependent DDE. In order to
apply a first order correction to the image
plane, we assume that
the direction dependent effects are constant enough across time, frequency,
and baseline. Then we can write:

$$[\mathcal{A}^H.\mathcal{A}](x,y) \sim N_{t,\nu,pq} p(x,y)\overline{\mathcal{D}^{H}\mathcal{D}}(x,y)$$


\noindent where $N_{t,\nu,pq}$ is the number if
baseline-time-frequency blocks, $\overline{\mathcal{D}^{H}\mathcal{D}}(x,y)$ is a $4\times4$ matrix being the average of
$\mathcal{D}^{t,\nu}_{pq}(y)^H.\mathcal{D}^{t,\nu}_{pq}(x)$ over
baselines, time, and frequency and
$p(x,y)$ is the PSF stacked in baselines, time, and frequency. If the
uv-coverage is good enough, then $\mathcal{A}^H.\mathcal{A}$ is block
diagonal ($p(x,y)=0$ for $x\neq y$, $p(x,y)=1$ otherwise). All
the $x\neq y$ terms cancel out in the final product, and in the
relevant part of $\overline{\mathcal{D}^{H}\mathcal{D}}$ matrix are the $N_{pix}$ $4\times4$ blocks
on the diagonal. Applying $\overline{\mathcal{D}^{H}\mathcal{D}}^{-1}$ to
$\mathcal{A}^H.V$ can then be done on each pixel separately, by
computing the product $\overline{\mathcal{D}^{H}\mathcal{D}}(x,x)^{-1}.I_x$ where
$I_x$ contains the full polarization information
$(XX_x,XY_x,YX_x,XY_x)$ for the $x^{th}$ pixel. This provides a way to
estimate an approximate least square clean component value from a flux
density in the dirty image in a given direction $\vec{s}_x$. The details
of this image plane normalization are further discussed in
Sec. \ref{sec:full_impl}. Although this normalization underlies a few
assumptions, from the simulations presented in
Sec. \ref{sec:convergence} we argue that the complex $4\times 4$ matrix
normalization per pixel presented here brings clear
improvement. However it does not seem
necessary for the sky solutions to convergence (see Sec. \ref{fig:dirtycorr}).

\begin{figure*}[ht!]
\begin{center}
\includegraphics[width=19cm]{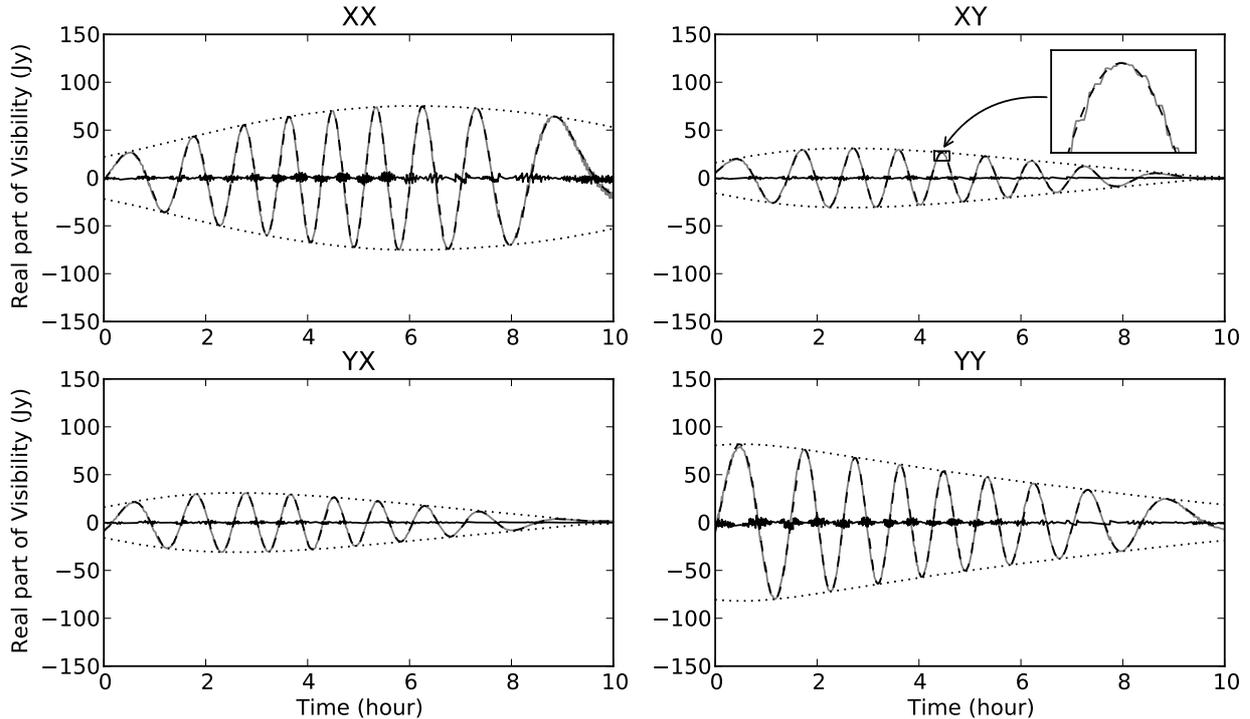}
\caption{\label{fig:degridding} The essential part of the A-Projection
  algorithm relies in the predict step, which transforms a 2D sky
  image (the projection of the celestial sphere on a plane) into a set
  of visibilities. We have simulated a dataset having one off axis
  source, where the {\it true} visibilities (black dashed line) have been estimated using
  Eq. \ref{eq:ME_Im} taking the beam into account. This plot shows the
  comparison in all measured polarizations between the exact value of
  the visibility of a given baseline, and the A-Projection estimate
  (gray line). Contrarily to a traditional predict step, the
  visibilities are modulated by the beam amplitude (dotted line) and we have
  time-dependent polarization leakage. The over-plotted graph shows a
  zoom in the small region shown on the top-right panel. In the
  degridding step, we use a computationally efficient closest
  neighbor interpolation, creating steps in the predicted
  visibilities. }
\end{center}
\end{figure*}


\section{Implementation of A-Projection for LOFAR}
\label{sec:impl}


As explained
above, the full polarization A-Projection scheme has been described in
\citet{Bhatnagar08}. However for the VLA implementation, due to the
approximate unitarity of VLA beams, only the diagonal Mueller matrix terms
had to be taken into account. As explained in Sec. \ref{sec:AProj},
LOFAR has got very wide fields of view, and the baseline-dependent Mueller matrices
associated to the 4-polarizations correlation products are
non-diagonal. This basically means that each individual polarization
cannot be treated independently from the others.
In this section we describe in detail a few implementations of
A-Projection allowing to correct for the non-diagonal Mueller
matrices. We propose optimizations in relation
to the telescope architecture. We will show in Sec. \ref{sec:sim} that
this algorithm can indeed deal with the heavily non-diagonal Mueller
matrices associated with the LOFAR's wide fields of view.

Following \citet{Bhatnagar08}, we have build our implementation of A-Projection on top of a tradition Cotton \&
Schwab CLEAN algorithm \citep{CS84}. It is an iterative deconvolution
algorithm that consists of two intricated steps.
The CLEAN and
A-Projection algorithms performs a series of operations that can be
described in the following way:

\begin{equation}
\label{eq:aproj_iter}
\begin{array}
{lcl}
\vspace{2mm}
\hat{I}^{n+1}&=&\hat{I}^n+\Phi.\mathcal{A}^H(V-\mathcal{A}\hat{I}^n)\\
\end{array}
\end{equation}

\noindent where the {\it true} sky estimate $\hat{I}^{n+1}$ at step
$n+1$ is built from its estimate $\hat{I}^{n}$ at step $n$ and $\Phi$
is a non-linear operator. It basically estimates the deconvolved sky
from the residual dirty image
$\mathcal{A}^H(V-\mathcal{A}\hat{I}^n)$. The minor cycle performs
operations that approximate $(\mathcal{A}^H\mathcal{A})^{-1}$
discussed in Sec. \ref{sec:AProj}, and takes the estimated image vector
to zero but at the strongest components above
a certain threshold. For the sky solutions to
converge, the predict step ($\mathcal{A}\hat{I}^n$ or {\it degridding}) has to be
accurate and unbiased. A-Projection is a fast way to apply
$\mathcal{A}$ or $\mathcal{A}^H$. If we separate the phase of the Fourier kernel in
Eq. \ref{eq:ME_Im} as the sum of $\phi^{0}(u,v,\vec{s})=u.l+v.m$ and
$\phi^{1}(w,\vec{s})=w.(\sqrt{1-l^2-m^2}-1)$, then folowing \citet{Bhatnagar08} invoking the
convolution theorem Eq. \ref{eq:ME_Im} becomes:

\begin{equation}
\label{eq:aproj}
\begin{array}
{lcl}
\vspace{2mm}
V^{t,\nu}_{pq}(u,v,w,i)&=& \mathcal{F}\left(\displaystyle\sum\limits^4_{j=1}\mathcal{D}_{pq}^{t,\nu}(i,j,\vec{s})W(w,\vec{s}) I_j(l,m)\right) \\ 
                     &=& \displaystyle\sum\limits^4_{j=1}\bigg[ \mathcal{F} \left( \mathcal{D}_{pq}^{t,\nu}(i,j,\vec{s}) W(w,\vec{s}) \right)\\
&& \ \ \ \ \ *\  \mathcal{F} \left( I_j\left( l,m \right) \right) \bigg]\\ 
\end{array}
\end{equation}

\noindent where $*$ is the convolution product, $\mathcal{F}$ is a 2D
Fourier transform, $W(w,\vec{s})=\exp{\left(-2 i\pi
  w.(\sqrt{1-l^2-m^2}-1)\right)}$, and $i$ and $j$ index the
polarization number (running over (XX, XY, YX, YY)).
This method is
efficient because the DDE are {\it smooth} on the sky, meaning the
support of the corresponding convolution function can be small (no
high frequency terms). Indeed, as shown in Fig. \ref{fig:beams} the
LOFAR station beam is very smooth, and depending on the field of view
the typical support is on the order of 5-11 pixels.

\subsection{Naive implementation}
\label{sec:full_impl}

The operation $\mathcal{A}\hat{I}$ in Eq. \ref{eq:aproj_iter} converts
a sky model into visibilities.  
To apply this operator on a massive amount of data
in an algorithmically efficient way, we apply the scheme outlined in
Eq. \ref{eq:aproj}. First (i) a 2-dimensional fast Fourier transform is applied
to the sky model image $\hat{I}$, on the 4 polarizations
independently. Then (ii) for each baseline in a
time-frequency block $\Delta(t,\nu)$ where the DDE is assumed to
be constant, the $16$ convolution functions
are computed as the Fourier transform of the image plane Kronecker product of
the DDE (for the LOFAR beam on a given baseline this block is
typically $10$ minutes and $0.2$ MHz wide). The
residual visibilities in each polarization are interpolated from the
sum in Eq. \ref{eq:aproj}. In practice, in order to minimize the support of
the W-term and associated aliasing effect, we multiply the DDE in the
image plane by a Prolate spheroidal function (see Appendix
\ref{sec:details} for more details).

The predicted visibilities
are removed from the measured visibility by computing the residual visibilities
$V_{\textrm{residual}}=V-\mathcal{A}\hat{I}^N$. Applying $\mathcal{A}^H$ applies a
correction to the residual visibilities, and projects the
result onto a grid (the {\it gridding} step) and Fourier
transforms it. In practice this is done as follows:

\begin{equation}
\begin{array}
{lcl}
V^{(t,\nu),(p,q)}_{\textrm{corr},u,v,w,i} &=&
\displaystyle\sum\limits^4_{j=1}\bigg(
\mathcal{F} \left( \mathcal{D}_{pq}^{*t,\nu}\left(j,i,\vec{s}\right) W^*\left(w,\vec{s}\right) \right)\\
&& \ \ \ \ \ *\   V^{(t,\nu),(p,q)}_{\textrm{residual},u,v,j}\bigg)\bigg]\bigg)\\ 
\end{array}
\end{equation}

\noindent and

\begin{equation}
\label{eq:aproj_resid}
\begin{array}
{lcl}
\vspace{2mm}
I_{\textrm{dirty}}(l,m,i) &=&
\mathcal{F}^{-1}\bigg(\displaystyle\sum\limits_{(t,\nu),(p,q)}V^{(t,\nu),(p,q)}_{\textrm{corr},u,v,w,i}\bigg)\\ 
\end{array}
\end{equation}

The resulting dirty image is still corrupted by the phased array beams
related effects
discussed in Sec. \ref{sec:poleffects}. Before the
minor cycle we can either multiply each
$x^{th}$ 4-polarization $I_{\textrm{dirty}}(x)$ pixel by the $4\times4$ matrix $\overline{\mathcal{D}^{H}\mathcal{D}}(x)^{-1}$, or simply normalize each
polarization of $I_{\textrm{dirty}}$ by $\overline{\mathcal{D}^{H}\mathcal{D}}_{ii}(x)$, the diagonal elements of $\overline{\mathcal{D}^{H}\mathcal{D}}(x)$.
As shown in Sec. \ref{sec:convergence}, fully applying
$\overline{\mathcal{D}^{H}\mathcal{D}}(x)^{-1}$ to each pixel in $I_{\textrm{dirty}}$
shows a minor improvement.


The computational cost of taking DDE into account using A-Projection
depends on (i) whether it is baseline dependent, (ii) the angular size
at which the effect needs to be sampled (thereby constraining the size
support of the convolution function) and (iii) the amplitude of the
terms of the $4\times4$ $\mathcal{D}^{t,\nu}_{pq}(\vec{s}_n)$ matrix. In the
case of the full polarization A-Projection, the data needs to be
corrected per baseline, per time and frequency slot. For each of those
data chunk, in order to recover the corrected 4-polarization
visibilities, one needs to take into account the 16 terms of the
$4\times4$ $\mathcal{D}^{t,\nu}_{pq}(x)$ Mueller matrix, and the 4
visibilities built from the 2D Fourier transform of the sky
model. Therefore in addition to the 16 convolution function estimate
per baseline and time-frequency slot, in the gridding and degridding
steps, one needs to compute $16\times N^2_S$
operations per 4-polarization visibility point, where $N_S$ is the
support of the convolution function. The algorithmic
complexity is further discussed in Sec. \ref{sec:cost}, but this
implementation is too slow to be practical.

\subsection{Separating antenna beam pattern and beam-former effect}
\label{sec:sep_element}

Depending on the assumptions and system architecture one can find
powerful algorithmic
optimizations.
We show here that in the case of LOFAR, we can use the fact that althought stations are
rotated one to another, the elementary antennas are parallel. The
effective phased array beam $B_{p,\vec{s}}$ of station $p$
is modeled as $B_{p,\vec{s}}=a_{p,\vec{s}}.E_{p,\vec{s}}$, where $a_{p,\vec{s}}$
is the array factor, and $E_{p,\vec{s}}$ is the {\it element beam}
pattern. The term $a_{p,\vec{s}}$ depends on the phased array geometry and on
the delays applied to the signal of the individual antennas before the
summation (by the {\it beam-former} of each individual LOFAR
stations). The term $E_{p,\vec{s}}$ models both the individual element
antenna sensitivity over the sky and its projection on the sky. We have

\begin{equation}
\label{eq:elem_eq}
\begin{array}
{lcl}
\textrm{Vec}(V^{corr}_{pq})    &=& \int\limits_{\vec{S}} \textrm{Vec}(a_{p,\vec{s}}.E_{p,\vec{s}} .I_{\vec{s}} .E^H_{q,\vec{s}}.a^*_{q,\vec{s}}) \\ 
&&.\exp{\left(-2 i\pi \phi(u,v,w,\vec{s})\right)} d\vec{s} \\ 
\end{array}
\end{equation}

\noindent with $a_{p,\vec{s}}$ being scalar valued and $E_{p,\vec{s}}$ is non-diagonal
because (intuitively) the element beam projection on the sphere depends on the
direction. Applying the convolution theorem to Eq. \ref{eq:elem_eq} we obtain:

\begin{equation}
\label{eq:sepel}
\begin{array}
{lcl}
\vspace{2mm}
\textrm{Vec}(V^{corr}_{pq})    &=& \mathcal{F}[E^*_{q,\vec{s}} \otimes
  E_{p,\vec{s}}]\\
\vspace{2mm}
&&*\ \mathcal{F}[a_{p,\vec{s}}.a^*_{q,\vec{s}}.\exp{\left(-2 i\pi \phi^{1}(w,\vec{s})\right)}] \\ 
&&*\int\limits_{\vec{S}} I_{\vec{s}} .\exp{\left(-2 i\pi \phi^{0}(u,v,\vec{s})\right)} d\vec{s} \\ 
\end{array}
\end{equation}

All LOFAR stations have different layout on the ground, so
the scalar valued array factor $a_{p,\vec{s}}$ is station
dependent. However all the individual antenna of all stations point at
the same direction and we can assume that the Mueller matrix is
baseline independent {\it ie} $E^*_{q,\vec{s}}
\otimes E_{p,\vec{s}}=E^*_{0,\vec{s}}
\otimes E_{0,\vec{s}}$. This requires an additional correction step of the gridded uv-plane
visibilities but as we will see below, this is an interesting
algorithmic shortcut because the element-beam correction can be
applied on the baseline-stacked grids.

The element beam is very smooth over the
sky and in most cases it can be assumed to be
constant at time-scales of an hour, so that the polarization
correction step does not need to be often applied. The degridding step
goes as follows: (i) in each time-frequency slot $\Delta(t,\nu)_E$
where the Mueller matrix of the element beam is assumed to be
constant, the polarization correction is applied to the (XX, XY, YX,
YY) grids as the sum of convolved grids by the terms of
$E^*_{0,\vec{s}} \otimes E_{0,\vec{s}}$. We then loop\footnote{We can
  parallelize the algorithm at this level.} over baseline
$(pq)$, and time-frequency range $\Delta(t,\nu)_a$ where the array
factor and w-coordinate are assumed to be constant within
$\Delta(t,\nu)_E$. For each step in the loop (ii) the oversampled
convolution function for baseline $(pq)$ is estimated in
$\Delta(t,\nu)_a$ for the term of the second line of
Eq. \ref{eq:sepel}, and (iii) it is used to interpolate the predicted
visibilities at the given uv-coordinate, separately on each polarization.

As explained in Sec. \ref{sec:cost}, the computing time for estimating
the convolution functions can be quite significant, and this scheme
allows us to compute one convolution function per baseline instead of
16, and 4 gridding/degridding steps instead of 16. We note however
that the assumption of baseline independence of the Mueller matrix on
which is based this optimization starts to be wrong in the cases of
direction dependent differential Faraday rotation, or for the longer baselines where
the curvature of the earth starts to have an importance (in that case
the element beam are not parallel). As discussed in
Sec. \ref{sec:cost}, this computing time of this implementation is
dominated by the convolution function estimate.



\subsection{Separating the W-term: hybrid w-stacking}
\label{sec:Wstack}

The support of the A-term is determined by the
minimum angular scale to be sampled in the image plane. The
beam or ionospheric effects are in general very smooth on the sky so
that little amount of pixels are needed to fully describe the effects,
therefore corresponding to a small convolution function support
size (typically $11\times11$ pixels). The highest spatial frequency in
the image plane is the W-term and its support can be as big as $\sim500\times500$
pixel for the long baselines, wide fields of view, when the target
field is at low elevation. This forces us to (i) compute a convolution
function with a large support, and (ii) grid each individual baseline
using the W-term dominated large convolution function.

We note however that the W-term is in itself baseline independent\footnote{\label{foot:bl} We talk about
baseline dependence when a set of baselines with exactly the same uvw coordinates
can give different visibilities.}: two
baselines characterized with different ionosphere, beams, but with the
same w-coordinate will have the same W-term. We therefore here
slightly change the piping of the algorithm by taking into account the
A-Term and the W-term separately as follows:

\begin{equation}
\label{eq:wstack}
\begin{array}
{lcl}
\vspace{2mm}
\textrm{Vec}(V^{corr}_{pq})    &=& \mathcal{F}[E^*_{\vec{s}} \otimes
  E_{\vec{s}}]\\
\vspace{2mm}
&&*\ \mathcal{F}[\exp{\left(-2 i\pi \phi^{1}(w_{plane},\vec{s})\right)}] \\ 
\vspace{2mm}
&&*\ \mathcal{F}[a_{p,\vec{s}}.a^*_{q,\vec{s}}.\exp{\left(-2 i\pi
    \phi^{1}(\Delta w,\vec{s})\right)}]\\
&&*\int\limits_{\vec{S}} I_{\vec{s}} .\exp{\left(-2 i\pi \phi^{0}(u,v,\vec{s})\right)} d\vec{s} \\ 
\end{array}
\end{equation}

We consecutively grid or degrid the data in w-slices {\it ie} that
have similar w-coordinates. This algorithm is also known as
W-stacking\footnote{See for example Maxim Voronkov's presentation at
  http://www.astron.nl/calim2010/presentations in the context of
  ASKAP}. In addition, we take into account the fact that the points
can lie above or below the associated w-plane central coordinate using
the term $\exp{\left(-2 i\pi \phi^{1}(\Delta w,\vec{s})\right)}$,
where $\Delta w=w-w_{plane}$. This step is similar to the traditional
w-projection algorithm.  If we have enough w-stacking planes, $\Delta
w$ is small, and the support of the baseline-time-frequency dependent
convolution function remains small, leading to a dramatic decrease of
the total convolution function estimation time. Conversely, given a
convolution function support we can find the maximum usable $\Delta w$
and derive the number of w-stacking planes as a function of the
observation's maximum $w$ coordinate (see Sec. \ref{sec:complWS} for
more detailed discussion). In the case of LOFAR, choosing a
convolution function support of $\sim21$ pixel gives a number of
w-stacking planes of $\sim30$.

This requires yet an additional step as compared to the implementation
described in Sec. \ref{sec:sep_element}. We describe below the
degridding step $\mathcal{A}\hat{I}^n$. First, following the notation introduced above (i) in
each time-frequency interval $\Delta(t,\nu)_E$, we correct the
4-polarization grids from the element beam (including projection
effects) using $E^*_{0,\vec{s}} \otimes E_{0,\vec{s}}$. Then (ii) we
loop over the number of w-planes (ranging from $-\mathrm{w}_{max}$ to
$\mathrm{w}_{max}$, see Sec. \ref{sec:wmax}), and convolve the grid
obtained in (i) by the associated w-term (appearing in the second line
of \ref{eq:wstack}). Finally in (iii) for each w-plane obtained in
each step of the loop (ii), we loop over the set of baselines
$(pq)_\mathrm{w}$ and time-frequency range
$\Delta(t,\nu)_\mathrm{a,w}$ associated with the given w-plane. We
interpolate the predicted visibilities at the given uv-coordinate,
separately on each polarization, based on the oversampled
$\mathcal{F}[a_{p,\vec{s}}.a^*_{q,\vec{s}}.\exp{\left(-2 i\pi
    \phi^{1}(\Delta w,\vec{s})\right)}]$ convolution function.
As discussed in Sec. \ref{sec:complWS}, this is the fastest
implementation of A-Projection we have obtained so far.

\section{Simulations}
\label{sec:sim}

In order to test the algorithmic framework described above we have
performed a series of tests on LOFAR simulated datasets. In this
section we summarize those results and discuss the computational costs
of A-Projection for LOFAR.

\subsection{One off-axis heavily polarized source}
\label{sec:on_source}

As discussed above by using A-Projection we can compute the {\it
  exact} value of the 4-polarization visibilities on a given baseline
at a given time and frequency, from (i) the sky model and (ii) the
baseline-time-frequency-direction dependent effects. In a first step, we focus
on testing the degridding full polarization A-Projection degridding
(or {\it predict}) step
($\mathcal{A}.\hat{I}$). The accuracy of this step is indeed vital for
the convergence of the CLEAN/A-Projection algorithm. Our experience
suggest that any small numerical systematic bias in this operation can
lead to strong divergence of CLEAN.

In order to test this transformation, we have simulated a dataset
having only one polarized off-axis point-like source in a low-band
$62$ MHz LBA dataset. The 4-polarization visibilities have been
generated using BBS (BlackBoard Selfcal\footnote{\label{foot:bbs} See
  http://www.lofar.org/wiki/doku.php?id=public:docu\-ments:lofar\_documents\&s[]=bbs
  for a review of BBS functionalities}), which computes a direct calculation of the visibilities
following Eq. \ref{eq:ME_Im}. It takes into account the beams of both
the individual elementary antenna (and their projection on the sky),
and the phasing of the individual antenna within a LOFAR station (the {\it array factor}). We
located the simulated source a few degrees from the phase center and
its flux density is non-physically polarized (with stokes parameters of
I,Q,U,V=100,40,20,10 Jy). Fig. \ref{fig:degridding} shows the real part
of BBS and A-Projection predicted visibilities on the baseline
$(01)$. The residuals are centered at zero, and A-Projection performs
very well in predicting the 4-polarization visibilities, taking into
account the complicated effects of the LOFAR phased array station's
beams. A traditional predict using simple Fourier transform, facets or
W-Projection would suffer from much higher residuals, driving
systematics in the deconvolution, thereby limiting the dynamic range. Here the residual errors are
dominated by the interpolation type we use in the uv domain (closest
neighborhood, see \ref{sec:details} for details).

\subsection{Dataset with many sources}

\subsubsection{LOFAR station beam}

In order to test our modified implementation of the whole CLEAN
algorithm (involving {\it gridding} and {\it degridding} steps), we
have simulated a dataset containing ~100 sources, with the source
count slope
following the $1.4$ GHz NVSS source count \citep{Condon98}. As in the
dataset described above, the visibilities are generated using
BBS$^{\ref{foot:bbs}}$. We have taken into account the Jones matrices
of the individual elementary antenna as well as the array-factor (the
beam-forming step). As explained in Sec. \ref{sec:sep_element} and
shown in Fig. \ref{fig:degridding}, as the LOFAR stations are rotated
one to another, all baselines will be effected by beam effects
differently. We have corrected the visibilities for the beam effect at
the first order by computing
$V^{t,\nu}_{pq,corr}=[D^{t,\nu}_p]^{-1}_{\vec{s}_0}V^{t,\nu}_{pq}[D^{t,\nu,H}_q]^{-1}_{\vec{s}_0}$,
where $D_p(\vec{s}_0)$ is the Jones matrix of station $p$ computed at
the center of the field. This mostly compensates for the element beam
effects, and more specifically the projection of the dipoles on the
sky. However, as shown below, the LOFAR fields are big, and the
projection of the dipoles vary across the field of view. Most of the
sources thereby generated have flux densities comprised between
$10^{-2}$ and $1$ Jy. We have added two bright sources of $100$ and
$10$ Jy.

\begin{figure*}[ht!]
\begin{center}
\hspace*{-1.3cm}
\includegraphics[width=20cm]{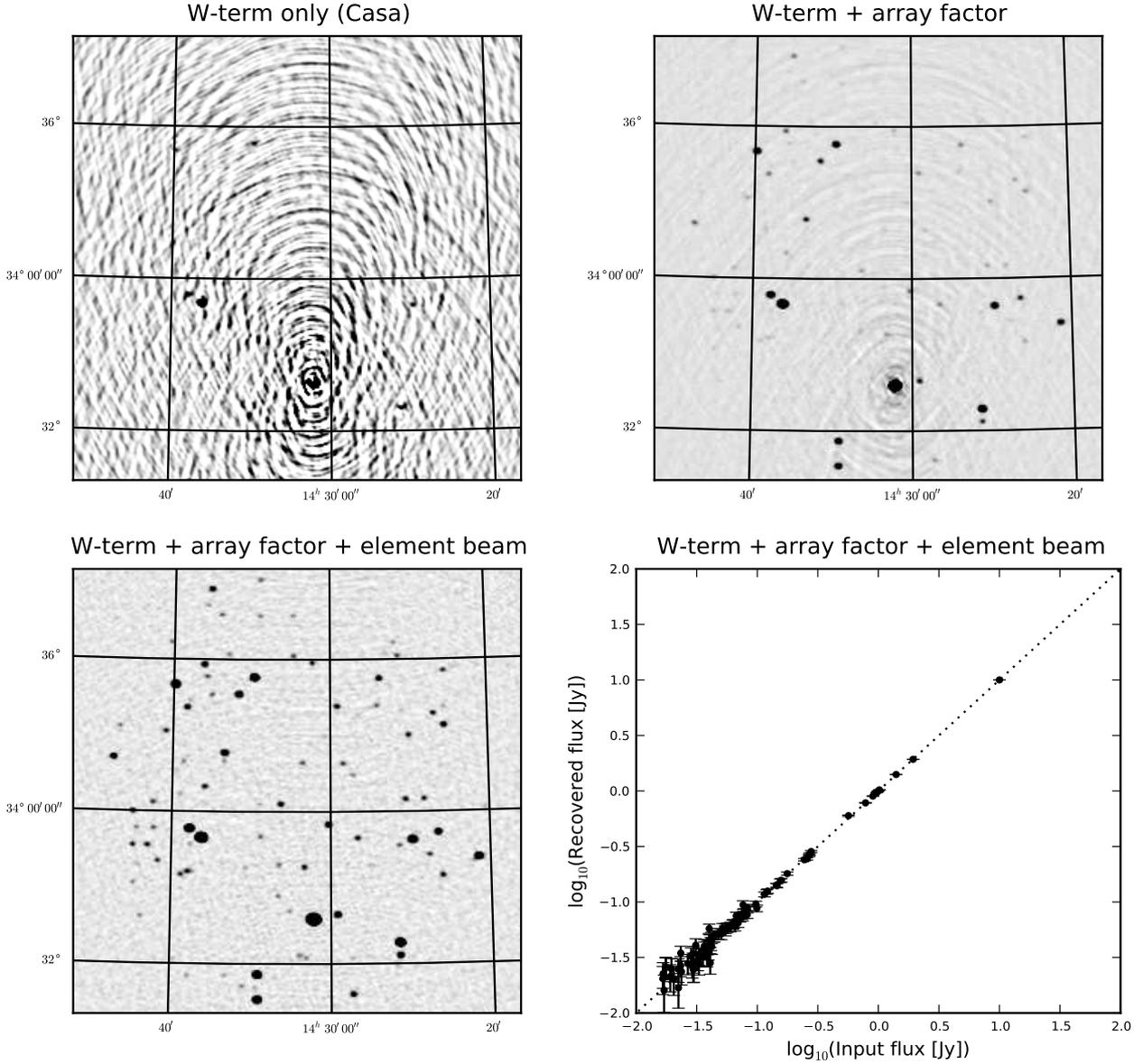}
\caption{\label{fig:AprojSimulIm} This figure shows the dramatic
  effect of the LOFAR phased array beams for a simulated
  dataset. Specifically, the visibilities have been generated taking
  into account (i) the individual antennas, (ii) their projection on
  the sky and (iii) the beam-forming step (the scalar {\it array
    factor}). The top-left image shows the deconvolved sky as
  estimated with a traditional imager not taking into account time-frequency
  direction dependent effects. The top-right and bottom-left have been
  generated by taking into account the array factor only and both array
  factor and the element beam respectively. The bottom right
  panel shows the input flux densities are correctly
  recovered. }
\end{center}
\end{figure*}

Fig. \ref{fig:AprojSimulIm} shows the restored images produced using
three different modified CLEAN algorithm. Each of those image contains
$\sim3000$ pixel in side, and is $\sim6$ degrees wide. We have used $15$
w-stacking, 128 $\Delta w$-planes (see Sec. \ref{sec:Wstack}), a maximum baseline of $5$ k$\lambda$, a Briggs\footnote{With
  a robust parameter of 0, see
  http://www.aoc.nrao.edu/dissertations/dbriggs} weighting
scheme, a cycle-factor of 2.5 and a number of minor cycle iterations of
10000. The first map has been
generated using W-Projection \citep{Cornwell08} as implemented in
CASA\footnote{http://casa.nrao.edu}. Strong artifacts are present around
the brightest off-axis source, and the dynamic range reaches $1:230$. In the second
image we have used our implementation of A-Projection taking into
account the array factor only. Taking that effect into account the
effect of the lower the residual visibility levels on each individual
baselines and the dynamic range reaches $\sim1:3.400$. In the third
image we have taken into account all the LOFAR beam effects: the
individual antenna sensitivity, their spatially varying projection, and the array
factor. The dynamic range increases to $\sim1:12.000$\footnote{It
  seems the output images of our
imager is presently limited at $\sim10^{-4}$ accuracy for some numerical
precision problem: all the code uses single precision floating point
numbers, and the roundings of products and sums involved in the
algorithm seems to generate limitations at this level. Detailed tests based on multi-threading and
single threading comparison consistantly reveal an instability at this
level. For the LOFAR surveys however, we might not need higher
accuracy, as we plan to use direction dependent pealing to substract
the brightest sources on the first $\sim1$:$10^2$ dynamic range.}

The sources flux densities in the restored maps have been
extracted using the LOFAR source extraction software pyBDSM\footnote{See
  http://www.lofar.org/wiki/doku.php for more information.}. As shown
in Fig. \ref{fig:AprojSimulIm}, the input flux densities are very well
recovered.

\subsubsection{LOFAR station beam and ionosphere}
\label{sec:ionosphere}

In order to test the ionospheric correction with A-Projection, we
have simulated a dataset containing 100 sources, affected by a simulated
ionosphere. The ionospheric effects essentially consist of a purely scalar, 
direction dependent phase. Faraday rotation is not included. In addition to this
purely scalar ionospheric phase effect the visibilities are affected by the
direction dependent LOFAR's stations beam effects discussed above.

The ionosphere is modeled as an infinitesimally thin layer at a height of 200
km. The Total Electron Content (TEC) values at a set of sample points are
generated by a vector autoregressive (var) random process. As described in
\citet{Tol09} the spatial correlation is given by Kolmogorov turbulence.
The values at intermediate  points are found using Kriging interpolation.
The set of sample points are the pierce points for five points in the image
located at the four corners and the centre. This way the ionospheric layer is
sampled in the most relevant area. There are at least five sample points within
the field of view of each station.

\begin{figure}[h!]
\begin{center}
\centering
\includegraphics[width=8cm]{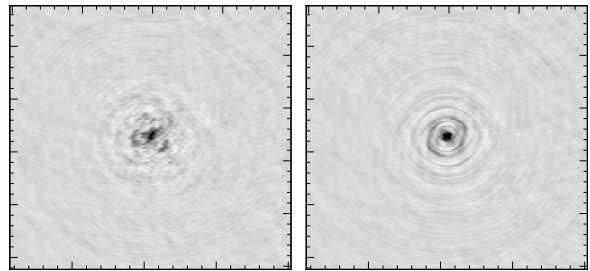}
\caption{\label{fig:ionpsf} This figure show the dirty image at the
  location of a bright source in a simulated dataset. The images are $1\deg$ in diameter. In the image of the left panel the dirty
  image shows important distortion as a result of ionospheric
  effect. The A-Projection correction (right panel) for an ionospheric phase screen
  clearly shows improvement. }
\end{center}
\end{figure}

\begin{figure*}[t!]
\begin{center}
\hspace*{-1.3cm}
\includegraphics[width=20cm]{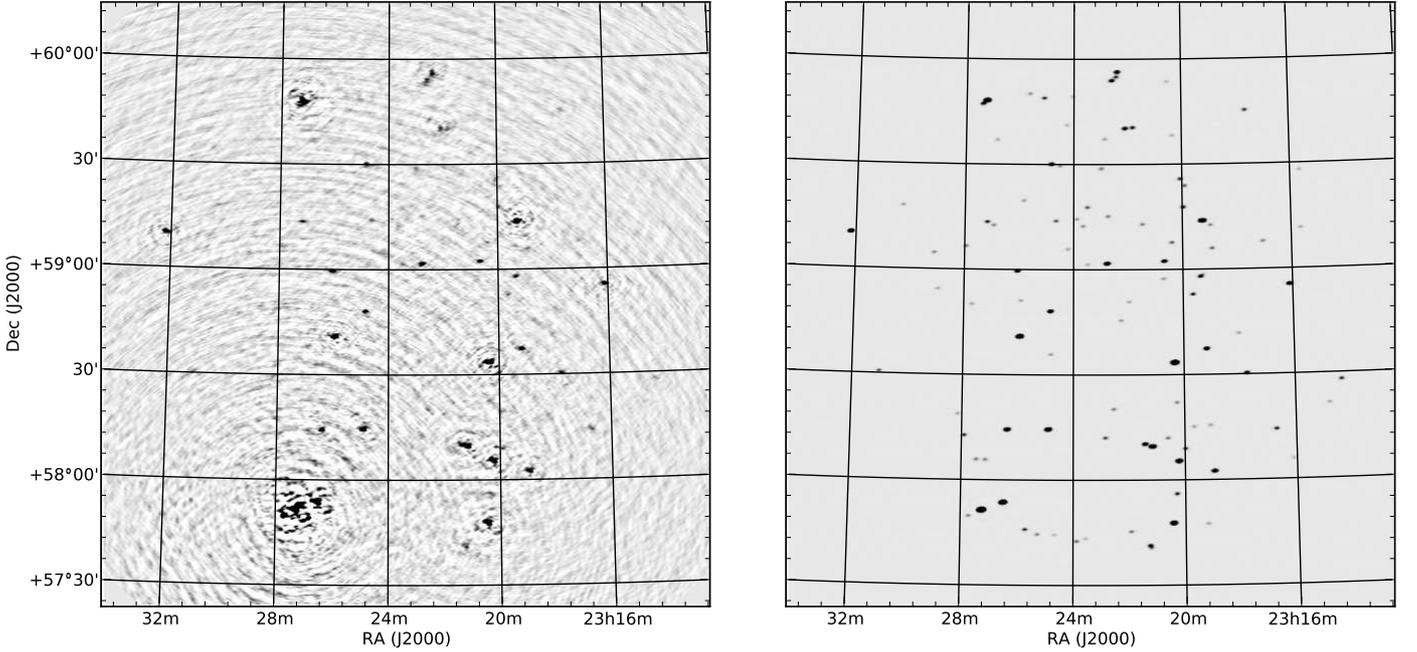}
\caption{\label{fig:ion} This figure shows the deconvolved image synthesized from
  the simulated dataset described in Sec. \ref{sec:ionosphere}. In the
  left panel, the ionospheric effects are not taken into account, and
  our deconvolution scheme naturally produces severe artifacts and high
  residuals in the reconstructed sky. The deconvolved image shown in the right
  panel has been estimated using our implementation of A-Projection
  with the time-dependent ionospheric phase screen.
}
\end{center}
\end{figure*}

Fig. \ref{fig:ionpsf} shows the dirty image at the location of a given
source before and after A-Projection correction of beam and
ionosphere. This suggets that the dirty undistorted sky is properly
recovered from the corrupted visibilities. We
compare in Fig. \ref{fig:ion} the cleaned image with and without
the A-Projection correction. Those simulations demonstrate that
A-Projection can indeed deal with the complicated effects associated
with the ionosphere.

\subsection{Convergence}
\label{sec:convergence}

We have also studied the influence of the various corrections
described in this paper on the convergence speed of the estimated flux
densities through the major cycle loop. For this we have used the
dataset described in Sec. \ref{sec:on_source} containing one off-axis
source having stokes (IQUV)=(100,40,20,10) Jy.

\begin{figure}[t!]
\begin{center}
\includegraphics[width=9cm]{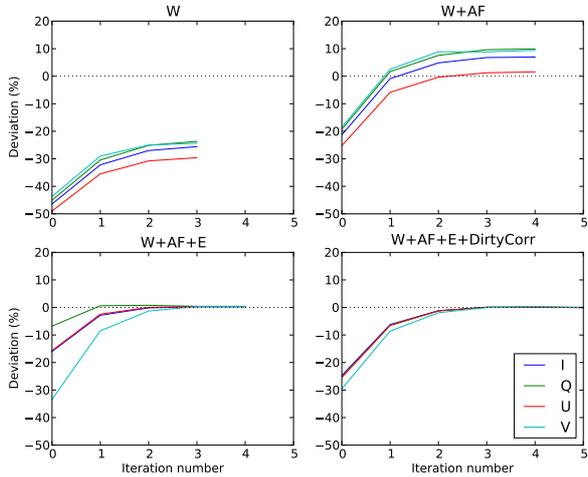}
\caption{\label{fig:dirtycorr} This figure shows the evolution of the
  estimated flux density as a function of the major cycle
  iteration for one polarized off-axis source. From top to bottom,
  left to right are generated (i) using W-Projection only, (ii) using
  the AW-Projection with the array factor only, (iii) using
  AW-Projection taking the full beam model into account. Finally in
  (iv) we show the effect of doing the image plane
  correction described in Sec.  \ref{sec:poleffects}.}
\end{center}
\end{figure}

Fig. \ref{fig:dirtycorr} shows the evolution of the estimated
flux density as a function of the major cycle iteration number for
different algorithm. In the first panel, the W-Projection algorithm
somewhat converges to the {\it observed} flux density (to the
"beam-multiplied" sky). The situation is better using full
polarization A-Projection without the element beam (therefore assuming
the dipole projection on the sky are constant across the field of
view). We note that in the absence of polarization, the situation is
not as bad. Taking the element beam into account, the algorithm makes
the estimated flux densities to converge to the true values to better
than a percent. In this version of the algorithm, the image plane
correction is the same on all polarization and is just a scalar
normalization (the average primary beam normalization). The situation
gets slightly better in terms of convergence speed by applying the image-plane renormalization
described in Sec. \ref{sec:poleffects}.

In any case the accuracy of the recovered flux densities seems to be
guaranteed by the accuracy of the degridding step. Our
experience in implementing A-Projection suggest that any small
systematic error in the degridding step can lead to biases in the
recovered flux densities or divergence in the more severe cases. The
image-plane polarization correction, seems to be bringing something
positive to the convergence speed, but that step appears not to be necessary.

\subsection{Computational and memory-related costs}
\label{sec:cost}

Given the large amount of data that have to be processed in the
imaging of a interferometric dataset, reducing the algorithmic
complexity is of primary importance.

\subsubsection{Memory-related issues}

The A-term is generally very smooth in the image plane, with
corresponding small support convolution functions and under those
conditions the A-Projection
algorithm is virtually free as explained in
\citet{Bhatnagar08} \citep[and in][in the case of
  W-Projection]{Cornwell08}: the A-term and the low w-coordinate
convolution function support is less or comparable to the spheroidal
function support that anyway needs to be applied in a traditional
gridder/degridder. However, as described in
Sec. \ref{sec:sep_element}, all LOFAR stations are rotated one to
another, and the synthesized stations beams are all different in a
given direction. This gives rise to a serious algorithmic complication
because the convolution functions become baseline-dependent: the
number of possible convolution functions is $16\times
N_{\mathrm{times}}\times N_{\mathrm{Freqs}}\times
N_{\mathrm{Stations}}\times (N_{\mathrm{Stations}}-1)/2$. With
$\sim800$ to $\sim1500$ baselines, even with a convolution function
every 10 minutes, a typical observing run of 8 hours, one frequency
channel, and an average support size of $\sim30$ pixel (taking into
account the w-term), this gives a $\sim100$ Gbytes of needed
storage. This is optimistic as in the case of other DDE such as the
ionosphere, the A-term will have to be evaluated every $30$
seconds. Even if the storage is done at the minimal necessary
resolution, those numbers are clearly prohibitive for storing the
convolution into memory. The convolution functions therefore have to
be computed on the fly, and algorithmically, this represents an
additional cost.

\subsubsection{Naive and element-beam-separated A-Projection}

\begin{figure}
\begin{center}
\includegraphics[width=9cm]{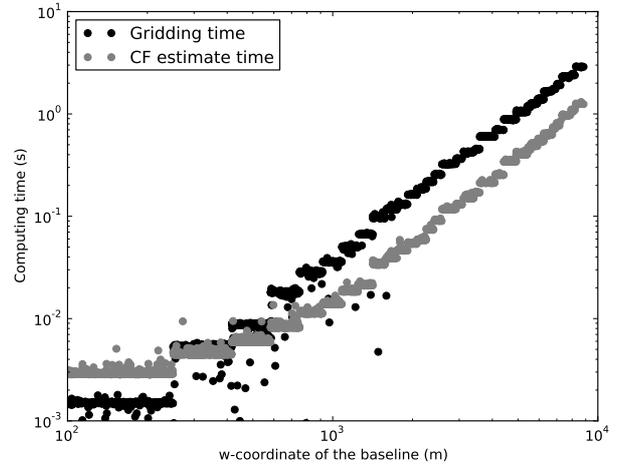}
\caption{\label{fig:times} Due to their different orientation, all
  LOFAR station have a different synthesized primary beam. The
  corresponding convolution function for applying the A-projection are
  therefore baseline-dependent, and we have to compute them on the
  fly. This figure is showing the computing time
  of both the gridding and convolution function estimate time for a 10
  degree field of view, a maximum w-coordinate to be $10^4$ meters,
  and a time-window of $T_{window}=1200$ seconds. When the DDE quickly
  varies, a convolution function is often required, and the
  convolution function computing time can largely exceed the gridding
  computing time.}
\end{center}
\end{figure}

The LOFAR station's beams are smooth on the sky, and the
corresponding convolution function support are small (typically
11-15 pixel complex image), while the W-term needs up to $\gtrsim$500
pixel depending on the w-coordinate (with an average of $\sim$30
pixels). The computing time depends on the convolution function
support $N_S$, and for the implementations described in
Sec. \ref{sec:full_impl} and \ref{sec:sep_element}, we have:

\begin{equation}
\label{eq:support}
\begin{array}
{lcl}
N_S&=&\displaystyle\sum\limits_{cf=\{S,A_p,A_q,W\}}N_S(cf)\\
\end{array}
\end{equation}

\noindent where the subscripts $S$, $A_p$, $A_q$, and $W$ stand for
the Prolate spheroidal ($\sim 7-11$ pixels), A-terms of antenna $p$
and $q$ and W-term. For a typical field of view, we have $N_S(A)=9-15$ and
$N_S(W)\appropto D^2.w$, where $D$ is the field of view diameter and $w$ is
the W-coordinate of the given baseline in the given time-slot (see
Appendix \ref{sec:details}). For that baseline, time and frequency
slot we can write the total computing
time as $t^{t\nu}_{pq,tot}=t^{t\nu}_{pq,grid}+t^{t\nu}_{pq,CF}$ where $t^{t\nu}_{grid}$ and $t^{t\nu}_{CF}$ are
the gridding and the convolution function estimate times. In most
cases for LOFAR data, we have $N_S\sim N_S(W)$ and $t_{grid}\appropto
D^4w^2$. Our experience has shown that $t_{CF}$ is dominated by the
computing time of the Fourier transform of the zero-padded convolution
function (see Fig. \ref{fig:times}), which size is $\mathcal{O}.N_S$ where $\mathcal{O}$ is the
oversampling parameter which controls the quality of the nearest
neighborhood interpolation. If $N^W_{buf}\propto1/(\Delta T_{win}\Delta \nu_{win})$ is the number of
visibility buffers associated with the W-plane ($\Delta T_{win}$ and
$\Delta \nu_{win}$ are the time/frequency window interval in which the
DDE are assumed to be constant), then we have
$t^W_{CF}\appropto N^W_{buf}.\mathcal{O}^2w^2D^4.\log(\mathcal{O}wD^2))$. The gridding time for a given w-plane is simply
$t^W_{grid}\propto N^W_{vis}w^2D^4$ where $N^W_{vis}$ is the number of
visibilities associated with the w-plane. The total computing time
can be written as:

\begin{equation}
\begin{array}
{lcl}
\vspace{0.2cm}
t_{tot}&\sim&N_{el}t_{el}+\bigg( \displaystyle\sum\limits_{W-planes}\bigg[at^W_{grid}+bt^W_{CF}\bigg]\bigg)\\
\end{array}
\end{equation}

\noindent where $a$ and $b$ are constants, $N_{el}$ is the number of
time/frequency blocks in which the element-beam is assumed to be
constant, and $t_{el}\appropto cN^2_{pix}(1+2\log(N_{pix}))$ is the
computing time necessary to apply the element-beam to the grids
($c=16$ for full polarisation imaging, $c=8$ for I-stokes only). For
the implementation described in Sec. \ref{sec:full_impl}, we have
$t_{el}=0$, but both $a$ and $b$ are $16$ times higher ($8$ for I-stokes only) than in the case of the
algorithm described in
Sec. \ref{sec:sep_element}. Fig. \ref{fig:times} shows the
gridding and convolution function estimate times as a function of the
w-coordinate of the given baseline in the given time-slots. From that
figure it is clear that (i) the W-term is the most important limiting
factor and (ii) the estimate of the convolution function
represents a major limitation of those implementations, especially in
the cases of a quickly varying DDE such as the ionosphere where the
convolution function calculation would largely dominate.

\subsubsection{Hybrid W-Stacking and A-Projection}
\label{sec:complWS}

As explained in Sec. \ref{sec:Wstack} for the W-staking
implementation, the baseline-time-frequency-dependent oversampled
convolution function is the Fourier transform of the zero-padded
image-plane product of the spheroidal, the A-term and a W-term
accounting for the $\Delta w$-distance between the given visibilities and the
corresponding w-plane. The bigger is the support of the convolution
function and the less w-planes we need to fully correct for the
w-term. As explained in Appendix \ref{sec:details}, in order to
properly sample the w-term in the image plane, to a given
w-coordinate and field of view correspond a convolution function
support. We can then obtain $\Delta w=a. N_S/ D^2$ (with $a=\sqrt{2}/(4\pi)$), and the
needed number of w-planes between $-w_{max}$ and $w_{max}$ is
$N_W= w_{max}D^2/(a. N_S)$. We compute the W-term convolution in
the image plane so this step cost goes as $t_W\propto N^2_{pix}(1+2\log(N_{pix}))$. The total computing time is then:

\begin{equation}
\begin{array}
{lcl}
\vspace{0.2cm}
t_{tot}
&\sim&bN_{vis}N_S^2+cN_{buf}\mathcal{O}^2N_S^2\log(\mathcal{O}N_S)\\
&&\;\;\;\;\;\;\;\;+N_{el}\big[t_{el}+N_Wt_W\big]\\


\end{array}
\end{equation}

\noindent where $N_{el}$ is the number of time/frequency buffers in
which the element-beam is assumed to be constant, $N_{buf}$ is the
number of time/frequency buffer where the A-/$\Delta w$-term are
assumed to be constant, $b$ and $c$ are
constants. In Tab. \ref{tab:Performance} we present the typical
computing times for a major cycle with the implementation discussed
here and presented in Sec. \ref{sec:Wstack}.

\begin{table*}
\caption[]{\label{tab:Performance} Computational performance of our fastest A-Projection implementation described in Sec. \ref{sec:Wstack} and \ref{sec:complWS}. 
Columns display the various performance corresponding to different
imaging settings. Those tests were performed for a 12 subbands dataset
(1 channel per subband), on the CEP2 LOFAR cluster node each havin 24
AMD Opteron(tm) 6172 Processors, and 64 Gb of RAM memory. $t^{grid}_{tot}$ is
the total for a gridding or degridding step. The times $t_{CF}$,
$t_{grid}$, $t_{el}$, $t_{W}$ are given in fraction of
$t^{grid}_{tot}$, and correspond to the times needed to compute
convolution functions, to gridding/degridding the data, to apply the element
beam, and to apply the W-term respectively. We are confident we can
still win a factor of $\gtrsim2-4$ with respect to those times values.}
\begin{center}
\begin{tabular}{|c|c|c|c|c|c|c|c|c||c|c|c|c|c|c|c|}
\hline
$N_{pix}$ & $d$        & D    & $N_{stokes}$ &$w_{max}$&$N_S$&$N_W$&$\Delta_t,\Delta_{\nu}$&$\Delta^{el}_t,\Delta^{el}_{\nu}$&$t_{CF}$&$t_{grid}$&$t_{el}$&$t_{W}$&Memory&$t^{grid}_{tot}$\\ 
         &($\arcsec$) & (deg)&             & (km)   &     &     &    (s, MHz)          &  (h, MHz)                   & (\%)   &(\%)     &(\%)    &(\%)    &(Gb)  & (sec.) \\ 
\hline
$1024$   & 40         & 11.4 & IQUV        & 20     & 15  & 14  & 300, 0.78            &             3, 3.5         &  67.8 &     10.2  &19.7  &  2.1  &  1   & 40.9 \\
$4096$   & 10         & 11.4 & IQUV        & 20     & 15  & 14  & 300, 0.78            &             3, 3.5         &  24.5 &     4.9   & 57.5 &  12.8 & 4.5  & 117.4\\
$4096$   & 10         & 11.4 & I           & 20     & 15  & 14  & 300, 0.78            &             none           &  52.0 &    11.4   & 0.0  &  36.5 & 1.5  & 143.7\\
$8192$   & 5          & 11.4 & I           & 20     & 15  & 14  & 300, 0.78            &             none           &  32.8 &     6.1   & 0.0  &  60.9 & 6.1  & 335.7\\

\hline
\end{tabular}
\end{center}
\end{table*}


\section{Summary and future work}
\label{sec:concl}

The new generation of low-frequency interferometers (EVLA, MWA, LWA)
and precursors of the SKA (LOFAR, ASKAP, MeerKAT) cannot be properly
used without the development of new techniques to calibrate and apply
in the imaging step the many direction dependent effects influencing
the electro-magnetic field. These effects mainly include the
antenna/stations complicated beam effects, ionosphere phase and
amplitude fast variation and associated Faraday rotation.

In this paper we have mainly discussed the issues associated with the
application of the LOFAR elementary antenna and station beams, which
involve the usage of a few levels of phase arrays. Using the
Measurement equation formalism, the associated high complexity (wide
field of view, individual station rotations, projection of the
elementary dipoles on the sky), the problem of imaging and
deconvolution of LOFAR calibrated dataset can be solved by applying
the A-Projection algorithm described in \citet{Bhatnagar08}. Due
to its very large field of view ($\sim5-10$ degrees in diameter), a
full-polarization A-Projection implementation dealing with
non-diagonal Mueller matrices is needed for LOFAR. In this paper we
have shown that A-Projection can indeed deal with the heavily
baseline-dependent non-diagonal direction-dependent Mueller matrices
associated with LOFAR baselines. We have also demonstrated that
effiscient ionospheric correction can be performed using
A-Projection. We have proposed a series of
implementations of A-Projection for LOFAR taking into account
non-diagonal Mueller matrices, aiming at accuracy and computing
efficiency.

\subsection{Optimizations}

As explained in Sec. \ref{sec:cost} the DDE although varying quickly are
smooth on the sky, so the convolution functions have a small
support. However, in the case of LOFAR the wide field of view imposes
to take into account (i) the W-term, and (ii) the off-diagonal terms
of the Mueller matrices (due to the varying projection of the dipoles
on the sky, or to the Faraday rotation). In addition (iii) the
baseline dependence make the storage of the convolution functions to
be prohibitive and those have to be computed on the fly.

In our first implementation (Sec. \ref{sec:full_impl}) the constraint
(i) makes the W-term convolution function support to often dominate,
while (ii) requires to set up a complicated machinery in the gridding
and degridding step, taking into account all polarizations to correct
for polarization leakage. In the case of a quickly varying DDE such as
the ionosphere, the step (iii) can completely dominate the computing
time (usually set by the gridding/degridding times).

Using the fact that LOFAR station's elementary antennas (responsible
for the complicated projection effects) are parallel although the
station's layout are rotated, we could separate the first
implementation in two steps (Sec. \ref{sec:sep_element}). The first is
a purely scalar gridding/degridding and suffers from (i) and (iii),
while the second is only affected by (ii). The non-diagonality of the
Mueller matrix is corrected on the baseline-stacked uv-plane,
so we win a factor between 10 and 16 as compared to the first
implementation. It is important to note that this optimization breaks
down when the non-diagonal Mueller matrix becomes baseline-dependent
such as in the cases of very long baselines (due to the earth's
curvature), or to the ionosphere's differential Faraday rotation.

In the last implementation (Sec. \ref{sec:Wstack}), we still apply the
direction-dependent non-diagonal Mueller matrix of the
baseline-independent element-beam, and go further by
separating the W-term from the A-Term. The former is responsible for the
large support sizes and is baseline independent (around a given
w-plane), while the later has small support and is baseline
dependent. In this implementation, we grid/degrid the data based on a
small support baseline-dependent oversampled convolution function, and
convolve the input or the output grids (forward and backward steps
respectively) with the non-oversampled W-term convolution function. We
save computing time by computing the oversampled convolution function
of a generally much smaller support, and the net gain lies between 2
and 10 as compared to the seconds implementation
(Sec. \ref{sec:sep_element}, and Tab. \ref{tab:Performance}).

Such optimizations are vital for the feasibility of the LOFAR
extragalactic surveys, given their huge integration times
totalizing hundreds of days. For the SKA, it will be important to take
the algorithmic optimization into account, as they are linked to the
instrument and system architecture. They can reduce the
algorithmic complexity by orders of magnitude.

\subsection{Real LOFAR data and future work}

We have conducted many experiments with LOFAR in order to test the
our imager on real LOFAR data. Specifically, we have observed the same
set of objects, in different observations having their pointing
centers shifted by few degrees. Compared to a traditional imager
(CASA), our implementation of A-Projection gives coherent corrected
flux densities at the level of 5-10\% on the edge of the
field. However, although we have shown in this paper that the
algorithm is giving excellent results on simulated dataset, it shows little
or no dynamic range improvement in the resulting images, as compared
to a traditional imager.

LOFAR is however a very complex machine and feeding the imager with a
wrong direction-dependent calibration solution will not lead to any
improvement in the deconvolved sky, and can of course even decrease the dynamical range. In order to improve LOFAR's
dynamic range we will have to make progress in understanding the
calibration aspects of DDE, especially those related to ionosphere and
differential Faraday rotation. Much effort is spent on that direction,
and DDE calibration algorithms of low complexity are under
development or have already been achieved such as SAGECal
\citep{Yatawatta08}.

On the imager side, further ongoing development include (i) implementation on GPU, of either
or both of the gridding/degridding or convolution function
calculation, (ii) compressed sensing in the image plane, (iii)
uv-plane interpolation techniques different from zero-padding FFT, and
(iv) Wide-Band A-Projection \citep{Bhatnagar12}. Ionosphere
and true beam calibration, in combination with pealing techniques,
will hopefully allow us to use the framework presented in this paper
to reach the high $\sim10^5-10^6$ dynamic range needed to construct
the deepest extragalactic LOFAR surveys, .

\begin{acknowledgements}
This work was done using the R\&D branch of the CASA code base. We would like
to thank the CASA Group and casacore team for the underlying
libraries. We thank Sarod
Yatawatta for the useful discussions on the element beam modeling that
lead to powerful optimizations.
\end{acknowledgements}

\bibliographystyle{aa}
\bibliography{references}


\begin{appendix}
\section{Measurement Equation formalism}
\label{sec:ME}

In order to model the complex direction dependent effects (DDE -
station beams, ionosphere, Faraday rotation, etc), we make extensive use of the
Measurement Equation formalism developed by \citet{Hamaker96}. The
Measurement Equation provides a model of a generic
interferometer. Each of the physical phenomena that transform or
convert the electric field before the correlation computed by the
correlator are modeled by linear transformations (2$\times$2
matrix). If $\vec{s}=(l,m,n=\sqrt{1-l^2-m^2})$ is a sky direction, and
$^{H}$ stands for the Hermitian transpose operator, then the
correlation matrix $V_{pq}$ between antennas $p$ and $q$, can be
written as follows:

\begin{equation}
\begin{array}
{lcl}
\vspace{1mm}
V^{meas}_{pq}  &=& G_p\left(\displaystyle\sum\limits_{\vec{s}}D_{p,\vec{s}}.K_{p,\vec{s}}.F_{\vec{s}}\ .\ F_{\vec{s}}^H.K^H_{q,\vec{s}}.D^H_{q,\vec{s}}\right)G^H_q \\ 
\end{array}
\end{equation}

\noindent where $D_{p,\vec{s}}$ is the product of direction-dependent
Jones matrices corresponding to antenna $p$ (e.g.,
beam, ionosphere phase screen, and Faraday rotation). $G_p$
is the product of direction-independent Jones matrices for antenna $p$
(like electronic gain and clock errors). The matrix
$K_{p,\vec{s}}$ describes the effect of the array geometry and
correlator on the observed phase shift of a coherent wavefront between
antenna $p$ and a reference antenna. This effect is purely scalar so
it is reducible to the product of a scalar and the unity matrix, so
we can write $K_{p,\vec{s}}.K^H_{q,\vec{s}}=\exp{\left(-2 i\pi
  \phi(u,v,w,\vec{s})\right)}.\vec{1}$
where$(u,v,w)$ is the baseline vector between antenna
$p$ and $q$ in wavelength units and $\vec{1}$ is the unity matrix. We then have
$\phi(u,v,w,\vec{s})=u.l+v.m+w.(\sqrt{1-l^2-m^2}-1)$, where the $-1$
term models the correlator effect when phasing the signals in the
direction of $w$. Finally
the product $F_{\vec{s}}\ .\ F_{\vec{s}}^H$ is the sky contribution in the
direction $\vec{s}$ and is the true underlying source coherency matrix
[[XX,XY], [YX, YY]].

This elegant formalism enables us to model the full polarization of
the visibility as a function of the {\it true} underlying electric
field correlation. In a simple and consistent way it takes the direction dependent and
direction independent effects into account. Indeed
most of the Jones matrices in the measurement equations have a fairly
simple formulation and radio calibration problem amounts to finding
the various components of $G$ and $E$, given a sky model
$I_{\vec{s}}=F_{\vec{s}}.F_{\vec{s}}^H$.

The measurement equation introduced above can be written in a more
extended and continuous form better suited for imaging by applying the
$\textrm{Vec}$ operator\footnote{\label{foot:vec} The $\textrm{Vec}$ operator
  transforms a matrix into a vector formed from the matrix columns
  being put on top of each other. It has the following useful properties:
  (i) $\textrm{Vec}(\lambda A)=\lambda \textrm{Vec}(A)$ ,(ii)
  $\textrm{Vec}(A+B)=\textrm{Vec}(A)+\textrm{Vec}(B)$, and (iii)
  $\textrm{Vec}(ABC)=(C^T \otimes A).\textrm{Vec}(B)$.} to
$V^{corr}_{pq}$. We obtain:

\begin{equation}
\label{eq:ME_Im0}
\begin{array}
{lcl}
\textrm{Vec}(V^{corr}_{pq})    &=&\int\limits_S (D^*_{q,\vec{s}} \otimes D_{p,\vec{s}}).\textrm{Vec}(I_{\vec{s}}) \\ 
 & & .\exp{\left(-2 i\pi \phi_{pq}(u,v,w,\vec{s})\right)} d\vec{s} \\ 
\end{array}
\end{equation}

\noindent where $\otimes$ is the Kronecker product, $\textrm{Vec}$ is
the operator that transforms a 2$\times$2 matrix into a dimension 4
vector.

\section{Further algorithm details}
\label{sec:details}

In this section we describe some important details of the
various implementations of A-Projection for LOFAR. In particular, we
make extensive use of the Prolate spheroidal function for resolution adaptation
and zero-padding for uv-plane interpolation.

As explained in Sec. \ref{sec:impl}, since LOFAR stations are
characterized by different primary beams, the gridding and degridding
steps are baseline dependent. Therefore the
convolution functions cannot be computed once and kept in memory as is done for W-projection. Instead, they have to be computed on the fly (see
Sec. \ref{sec:cost}). However, because the station's beam model is fairly
complex and costly to evaluate (coordinates transformation, estimate of
the Element beam Jones matrix), we store in memory at the minimal
resolution the 4-polarization image plane beams (their Jones
matrices). The necessary resolution is simply estimated as
$\lambda/(2.D_{station})$, where $D_{station}$ is the given station's
diameter.

The W-term is also estimated once and stored in memory at
the minimal resolution. This amounts to finding the maximum frequency to
be sampled in the image plane, and the corresponding number of pixels
corresponding to the minimum support required for the W-term convolution
function. If the image is of angular diameter $D_{im}$, the necessary
resolution needed to properly sample to W-term is the inverse of the
highest spatial frequency, located in one of its corners. The support
of the W-term is then $N_W=(4\pi w D_{im}^2)/\sqrt{2-D_{im}^2}\sim 4\pi w D_{im}^2/\sqrt{2}$.

In order to interpolate the visibilities on the grid in the gridding
step (or conversely in the degridding step), or to adapt the
resolution of the A- and W-terms we use a zero-padding
interpolation. This scheme can produce artifacts due to the presence
of sharp edges and aliasing
problems, so we have to make extensive use of a Prolate spheroidal
function. It is computed at the maximum resolution in the image
plane. We then Fourier transform it, find its
support $N_S(\mathrm{S}^{uv}_{ph})$, "cut" it to that size, and store
it in the uv-plane (hereafter $\mathrm{S}^{uv}_p$).

For the various algorithms presented in this paper, we have to compute
the products of various DDE in the image plane. For example for the
algorithm described in Sec. \ref{sec:full_impl}, we have to
adapt the A- and W-terms resolution before multiplying them in the
image plane. We first have to find
the support $N_S$ of the convolution function as in Eq. \ref{eq:support}. We first compute
the image plane spheroidal at the resolutions of the A- and W-terms
($\mathrm{S}_p^{N_S(A)}$ and $\mathrm{S}_p^{N_S(W)}$ respectively)
as follows:

\begin{equation}
\begin{array}
{ccl}
\vspace{2mm}
\mathrm{S}_p^{N_S(A)}&=&\mathcal{F}^{-1}\mathcal{Z}^{N_S(A)}\mathrm{S}^{uv}_p\\
\vspace{2mm}
\mathrm{S}_p^{N_S(W)}&=&\mathcal{F}^{-1}\mathcal{Z}^{N_S(W)}\mathrm{S}^{uv}_p\\
\end{array}
\end{equation}

\noindent where $\mathcal{F}$ is the Fourier transform, $\mathcal{Z}^{n}$ is the zero-padding operator
that puts the input into a grid of the size $n$. To estimate the
A-term interpolated on an $N_S\times N_S$ pixel image:

\begin{equation}
\begin{array}
{ccl}
\mathrm{A}^{N_S}&=&(\mathrm{S}^{N_S})^{-1}\mathcal{F}^{-1}\left(\mathcal{Z}^{N_S}.\mathcal{F}\left(\mathrm{S}_p^{N_S(A)}\mathrm{A}\right)\right)\\
\end{array}
\end{equation}

We obtain the image plane effects at the same resolution and multiply them according to the
specific needs of the various implementations described in
Sec. \ref{sec:impl} and obtain the image plane product $P_{im}$. We
then compute the oversampled convolution
function as:

\begin{equation}
\begin{array}
{ccl}
\vspace{2mm}
\mathrm{CF}^{\mathcal{O}N_S}&=&\mathcal{F}\left(\mathcal{Z}^{\mathcal{O}N_S}\left(\mathrm{S}^{N_S}P_{im}\right)\right)\\
\end{array}
\end{equation}

\noindent where the resulting interpolated convolution function
$\mathrm{CF}^{\mathcal{O}N_S}$ has $\mathcal{O}N_S\times
\mathcal{O}N_S$ pixels with $\mathcal{O}$ the oversampling parameter. If the final image is of
size $N_{pix}\times N_{pix}$, because we have used the
spheroidal function, after applying the inverse Fourier transform, we
have to normalise the dirty image by the spheroidal function
$\mathrm{S}_p^{N_{pix}}=\mathcal{F}^{-1}\mathcal{Z}^{N_{pix}}\mathrm{S}^{uv}_p$.

As outlined above, all LOFAR stations are different, and the
convolution functions are baseline dependent. The {\it for} loops described in
Sec. \ref{sec:impl} are therefore parallelized on baseline. For the
degridding step from a common read-only grid, the residual
visibilities are estimated independently using different threads.
The code has been created in the LOFAR package and is
is dependent on the casacore and casarest packages.

\section{Limiting the maximum w-coordinate}
\label{sec:wmax}

For the traditional interferometers at high frequencies, in general
the field of view is small enough so that the W-term can always be
neglected ($w.(\sqrt{1-l^2-m^2}-1)\sim 0$). However, for the wide
fields of view, long baselines interferometers, the W-term is of great
importance, and not taking it into account produces artifacts and
considerably reduces the dynamic range. For a given field of view, or
a given angular distance to the phase center, the importance of the
W-term increases as the w-coordinate value, {\it i.e.} when the targeted
field is at low elevation. It is therefore important to stress that wide
fields of view or long baselines do not directly mean that the W-term will take an
importance: whatever the baseline or field of view, a planar array
that would observe at the zenith would always give a null w-coordinate.

\begin{figure}[h!]
\begin{center}
\hspace{-0.5cm}
\includegraphics[width=9.5cm]{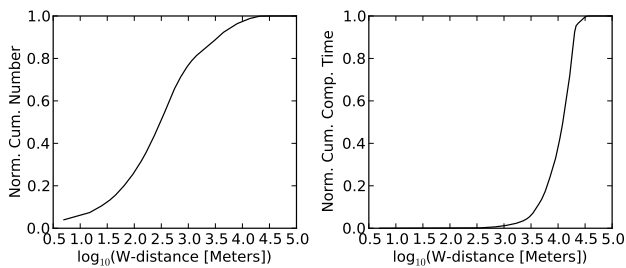}
\caption{\label{fig:w-term} For a given field of view the W-term
  increases for lower elevation of the target field. Using the W-Projection
  algorithm, the computing time increases with $w^2$. This figure
  shows the normalized cumulative distribution of the w-coordinate (left panel) for a
  typical LOFAR observation, and the corresponding normalized
  cumulative computing time (right panel). We can see that rejecting
  the $\sim5\%$ of the points with $w>10^4$ saves $\sim70\%$ of the
  computing time.}
\end{center}
\end{figure}

Algorithmically, for A-Projection, the W-term is one of the main
limiting factor. Using the W-Projection algorithm \citep{Cornwell08},
assuming the W-term support is higher that the Prolate spheroidal
support, the gridding time evolves as $t_{\mathrm{grid},w}\propto
w^2.D^4$, because the highest spatial frequency in the image plane has
to be properly sampled. We found that on a typical LOFAR dataset
this non-linear behavior generally makes the $\lesssim5\%$ of the
points with the highest w-coordinates to be responsible for
$\gtrsim70\%$ of the computing time (as in
Fig. \ref{fig:w-term}). This little amount of data does not necessarily
bring sensitivity or resolution. Setting a {\it wmax} value above
which the visibilities are not gridded significantly increases the
computational efficiency, without loosing sensitivity or resolution.

\end{appendix}

\end{document}